\documentclass{pnastwo}

\usepackage{mathrsfs}  
\usepackage{mathtools}
\usepackage{amssymb}
\usepackage{amsmath}
\usepackage{algorithm}
\usepackage[noend]{algpseudocode}
\usepackage{float}
\usepackage{cite}

\begin{document}

\title{Solving ptychography with a convex relaxation}

\author{Roarke Horstmeyer\affil{1}{Department of Electrical Engineering, California Institute of Technology, Pasadena, CA
91125},
Richard Y. Chen\affil{2}{Department of Computing and Mathematical Sciences, California Institute of Technology, Pasadena, CA
91125},
Xiaoze Ou\affil{1}{},
Brendan Ames\affil{3}{Department of Mathematics, University of Alabama, Tuscaloosa, AL 35487},
Joel A. Tropp\affil{2}{},
\and
Changhuei Yang\affil{1}{}}

\date{}


\maketitle

\begin{article}
\begin{abstract}
{Ptychography is a powerful computational imaging technique that transforms a collection of low-resolution images into a high-resolution sample reconstruction. Unfortunately, algorithms that are currently used to solve this reconstruction problem lack stability, robustness, and theoretical guarantees. Recently, convex optimization algorithms have improved the accuracy and reliability of several related reconstruction efforts. This paper proposes a convex formulation of the ptychography problem. This formulation has no local minima, it can be solved using a wide range of algorithms, it can incorporate appropriate noise models, and it can include multiple a priori constraints. The paper considers a specific algorithm, based on low-rank factorization, whose runtime and memory usage are near-linear in the size of the output image. Experiments demonstrate that this approach offers a 25\% lower background variance on average than alternating projections, the current standard algorithm for ptychographic reconstruction.}
\end{abstract}

\keywords{ptychography | Fourier ptychography | convex optimization | phase retrieval | structured illumination | synthetic aperture imaging}


\section{Introduction}

Over the past two decades, ptychography~\cite{Nellist95, Nugent10} has surpassed all other imaging techniques in its ability to produce high-resolution, wide field-of-view measurements of microscopic and nanoscopic phenomena. Whether in the X-ray regime at third-generation synchrotron sources~\cite{Rodenburg07,Thibault08,Dierolf10,Shapiro14}, in the electron microscope for atomic scale phenomena~\cite{Hue10}, or in the optical regime for biological specimens~\cite{Maiden10}, ptychography has shown an unparalleled ability to acquire hundreds of megapixels of sample information near the diffraction limit. The standard ptychography principle is simple: a series of diffraction patterns are recorded from a sample as it is scanned through a focused beam. These intensity-only measurements are then computationally converted into a reconstruction of the complex sample (i.e., its amplitude and phase), which contains more pixels than a single recorded diffraction pattern.

A recently introduced imaging procedure, termed Fourier ptychography (FP), uses a similar principle to create gigapixel optical images with a conventional microscope~\cite{Zheng13}. The only required hardware modification is an LED array, which illuminates a stationary sample from different directions as the microscope captures a sequence of images. As in standard ptychography, FP must also recover the sample's phase as it merges together the captured image sequence into a high-resolution output. Standard and Fourier ptychographic data are connected via a linear transformation~\cite{Horstmeyer14a}, which allows both setups to use nearly identical image reconstruction algorithms.  

Standard and Fourier ptychography both avoid the need for a large, well-corrected lens to image at the diffraction-limit. Instead, they shift the majority of resolution-limiting factors into the computational realm. Unfortunately, an accurate and reliable solver does not yet exist. For ptychography to succeed, we must reconstruct the phase of the scattered field, which is a non-convex problem. To date, nearly all ptychography algorithms solve the phase retrieval problem with an iterative technique called alternating projections (AP)~\cite{Gerchberg72,Fienup82}. However, AP and simple variants~\cite{Elser07,Marchesini07} often converge to incorrect local minima or they stagnate~\cite{Fienup86}. Few guarantees exist regarding convergence, let alone convergence to a reasonable solution. Despite these shortcomings, many authors have pushed beyond the basic algorithms~\cite{Faulkner04} to account for unknown system parameters~\cite{Maiden09, Maiden13}, to improve outcomes by careful initialization~\cite{Marchesini13}, and to enable multiplexed acquisition~\cite{Tian14}.

In this article, we formulate a convex program for the ptychography problem, which allows us to use efficient computational methods to obtain a reliable image reconstruction. Convex optimization has recently matured into a powerful computational tool that now solves a variety of challenging problems~\cite{Boyd}. However, many sub-disciplines of imaging, especially those involving phase retrieval, have been slow to reap its transformative benefits. Several prior works~\cite{Bauschke02, Fazel02,Balan09,Shechtman11,Candes12} have connected convex optimization with abstract phase retrieval problems, but this is the first work that applies convex optimization to the quickly growing field of high-resolution ptychography.

The convex optimization approach to ptychographic reconstruction has many advantages over AP.  Our formulation has no local minima, so we can always obtain a solution with minimum cost. The methodology is significantly more noise-tolerant than AP, and the results are more reproducible. There are also opportunities to establish theoretical guarantees using machinery from convex analysis.

Furthermore, there are many efficient algorithms for our convex formulation of the ptychography problem. To obtain solutions at scale, we apply a factorization method due to Burer and Monteiro~\cite{Burer03,Burer05}. This method avoids the limitations of earlier convex algorithms for abstract phase retrieval, whose storage and complexity scale cubically in the number of reconstructed pixels~\cite{Candes12}. Moreover, recent results establish that this factorization technique converges globally under certain conditions~\cite{Sa14}, offering a promising theoretical guarantee. The end result is a new, noise-tolerant algorithm for ptychographic reconstruction that is efficient enough to process the multi-gigapixel images that future applications will demand. 

Here is an outline for the paper. First, we develop a linear algebraic framework to illustrate the Fourier ptychographic image formation process. Second, we manipulate this framework to pose its sample recovery problem as a convex program. This initial algorithm, termed ``convex lifted ptychography" (CLP), supports a-priori knowledge of noise statistics to significantly increase the accuracy of image reconstruction in the presence of noise. Third, we build upon research in low-rank semidefinite programming~\cite{Burer03,Burer05} to develop a second non-convex algorithm, called ``low-rank ptychography" (LRP), which improves on the computational profile of CLP. Finally, we explore the performance of LRP in both simulation and experiment to demonstrate how it may be of great use in reducing the image capture time of Fourier ptychography.

\begin{figure}
\centerline{\includegraphics[width=.48\textwidth]{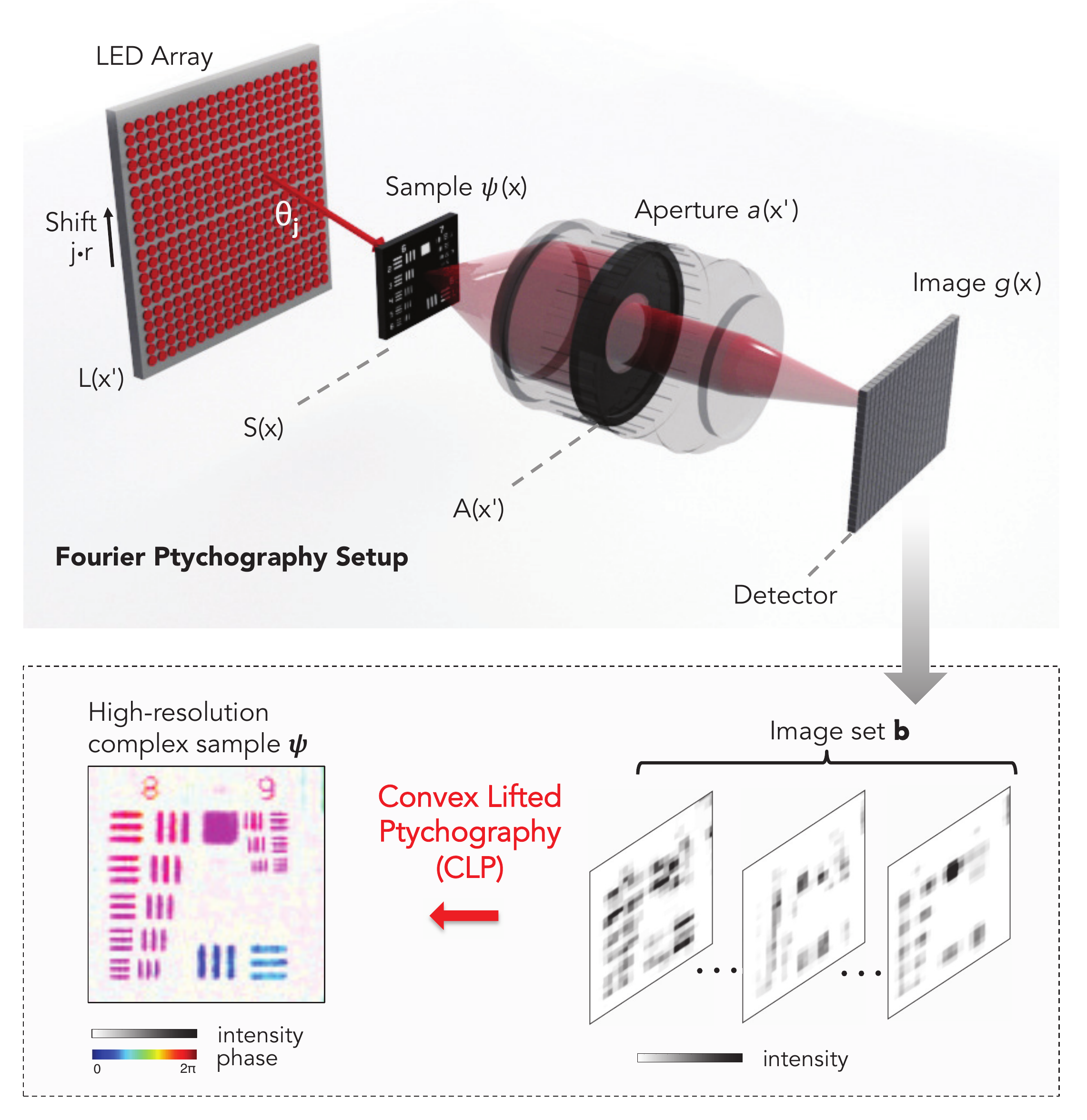}}
\caption{Diagram of the Fourier ptychography setup (top), where we use an LED array to illuminate a sample from different directions and acquire an image set $\mathbf{b}$ (bottom). This paper introduces a convex phase retrieval algorithm to transform this image set into a high-resolution complex sample estimate $\psi$. Included image set and  reconstructed resolution target are experimental results. \label{setup}}
\end{figure}

\section{Optical setup}

In this section, we outline the image capture process for Fourier ptychography (see Fig.\ \ref{setup}). At the end of this section, we discuss how a simple exchange of variables yields a nearly equivalent mathematical description of ``standard" (i.e., diffraction imaging-based) ptychography data, which our proposed algorithm may also process. In addition, while the following analysis considers a two-dimensional optical geometry for simplicity, extension to three dimensions is direct.

We assume that a distant plane $L(x')$ contains $q$ different quasi-monochromatic LEDs (central wavelength $\lambda$) evenly distributed along $x'$ with a spacing $r$. We assume each LED acts as an effective point source and illuminates a sample $\psi(x)$ at a plane $S(x)$ a distance $l$ away from $L(x')$. Under this assumption, the $j$th LED illuminates the sample with a spatially coherent plane wave at angle $\theta_{j}$ = $\tan^{-1}{(jr/l)}$, where $-q/2\le j\le q/2$. Additionally assuming the sample $\psi(x)$ is thin, we may express the optical field exiting the thin sample as the product,
\begin{equation}
s(x, j) = \psi(x)e^{ikxp_{j}},
\label{sample}
\end{equation}
where the wavenumber $k=2\pi/\lambda$ and $p_{j}=\sin{\theta_{j}}$ describes the off-axis angle of the $j$th LED. The $j$th illuminated sample field $s(x,j)$ then enters an imaging system with a low numerical aperture (NA).  Neglecting scaling factors and a quadratic phase factor for simplicity, Fourier optics gives the field at the imaging system pupil plane $A(x')$ as $\mathcal{F}\left[s(x,j)\right]=\hat{\psi}(x'-p_{j})$, where $\mathcal{F}$ represents the Fourier transform between conjugate variables $x$ and $x'$, $\hat{\psi}$ is the Fourier transform of $\psi$, and we have applied the Fourier shift property. The shifted spectrum field $\hat{\psi}(x'-p_j)$ is then modulated by the imaging system's aperture function $a(x')$, which acts as a low-pass filter. It is now useful to consider the spectrum $\hat{\psi}$ discretized into $n$ pixels with a maximum spatial frequency $k$. We denote the bandpass cutoff of the aperture function $a$ as $k \cdot m/n$, where $m$ is an integer less than $n$. The modulation of $\hat{\psi}$ by $a$ results in a field characterized by $m$ discrete samples, which propagates to the camera imaging plane and is critically sampled by an $m$-pixel digital detector. This forms a reduced-resolution image, $g$:
\begin{equation}
g(x, j) = \left|\mathcal{F}\left[a(x')\hat{\psi}(x'-p_{j})\right]\right|^{2}.
\label{image}
\end{equation}
$g(x, j)$ is an $(m \times q)$ Fourier ptychography data matrix. Its $j$th column contains a low-resolution image of the sample while it is under illumination from the $j$th LED.

The goal of Fourier ptychographic post-processing is to reconstruct a high-resolution ($n$-pixel) complex spectrum $\hat{\psi}(x')$, from the multiple low-resolution ($m$-pixel) intensity measurements in the data matrix $g$. Once $\hat{\psi}$ is found, an inverse-Fourier transform will yield the desired complex sample reconstruction $\psi$. As noted above, most current ptychography setups solve this inverse problem using alternating projections (AP): after initiating a complex sample estimate $\psi_{0}$, iterative constraints help force $\psi_{0}$ to obey all known physical conditions. First, its amplitude is forced to obey the measured intensity set from the detector plane (i.e., the values in $g$). Second, its spectrum $\hat{\psi}_{0}$ is forced to lie within a known support in the plane that is Fourier conjugate to the detector. While such a projection strategy is known to converge when each constraint set is convex, the intensity constraint applied at the detector plane is not convex~\cite{Bauschke03}, leading to erroneous solutions~\cite{Sanz84} and possible stagnation~\cite{Fienup86}.  

The Fourier ptychography setup in Fig.\ \ref{setup} may be converted into a standard ptychography experiment by interchanging the sample plane $S$ and the aperture plane $A$. This results in a standard ptychographic data matrix taking the form of Eq.\ \ref{image} but now with a sample spectrum described in real space as $\psi$, which is filtered by the Fourier transform of the aperture function, $\hat{a}$. These two simple functional transformations lead to a linear relationship between standard and Fourier ptychographic data~\cite{Horstmeyer14a}. To apply the algorithmic tools outlined next to standard ptychography, simply adhere to the following protocol wherever either variable appears: 1) replace the sample spectrum $\hat{\psi}$ with the sample function $\psi$, and 2) replace the aperture function $a$ with the shape of the focused field that illuminates the sample, $\hat{a}$, in standard ptychography setups.

 \section{Convex Lifted Ptychography}
 
We begin the process of solving Eq.\ \ref{image} as a convex program by expressing it in matrix form. First, we represent the unknown sample spectrum $\hat{\psi}$ as an $(n \times1)$ vector. Again, $n$ is the known sample resolution before it is reduced by the finite bandpass of the lens aperture. Second, the $j$th detected image becomes an $(m \times 1)$ vector $\mathbf{g}_j$, where again $m$ is the number of pixels in each low-resolution image. The ratio $n/m$ defines the ptychographic resolution improvement factor, set by the ratio between the maximum LED angle and the imaging lens acceptance angle. 

Third, we express each lens aperture function $a(x+p_j)$ as an $(n \times1)$ discrete aperture vector $\mathbf{a}_j$, which modulates the unknown sample spectrum $\hat{\psi}$. To rewrite Eq.\ \ref{image} as a matrix product, we define $\{\mathbf{A}_j\}_{j=1}^q$ to be the sequence of $(m \times n)$ rectangular matrices that contain a deterministic aperture function $\mathbf{a}_j$ along a diagonal. For an aberration-free rectangular aperture, each matrix $\mathbf{A}_j$ has ones originating at $(0, p'_j)$ and terminating at $(m, p'_{j}+m-1)$, where $p'_j$ is now a discretized version of our shift variable $p_j$. Finally, we introduce an $m\times m$ discrete Fourier transform matrix $\mathbf{F}^{(m)}$ to express the transformation of the low-pass filtered sample spectrum through our fixed imaging system for each low-resolution image $\mathbf{g}_j$:
\begin{equation}
\mathbf{g}_{j} = \left|\mathbf{F}^{(m)}\mathbf{A}_{j}\hat{\psi}\right|^{2}, \quad 1\leq j \leq q.
\label{image_matrix_2}
\end{equation}

\begin{figure}
\centerline{\includegraphics[width=.48\textwidth]{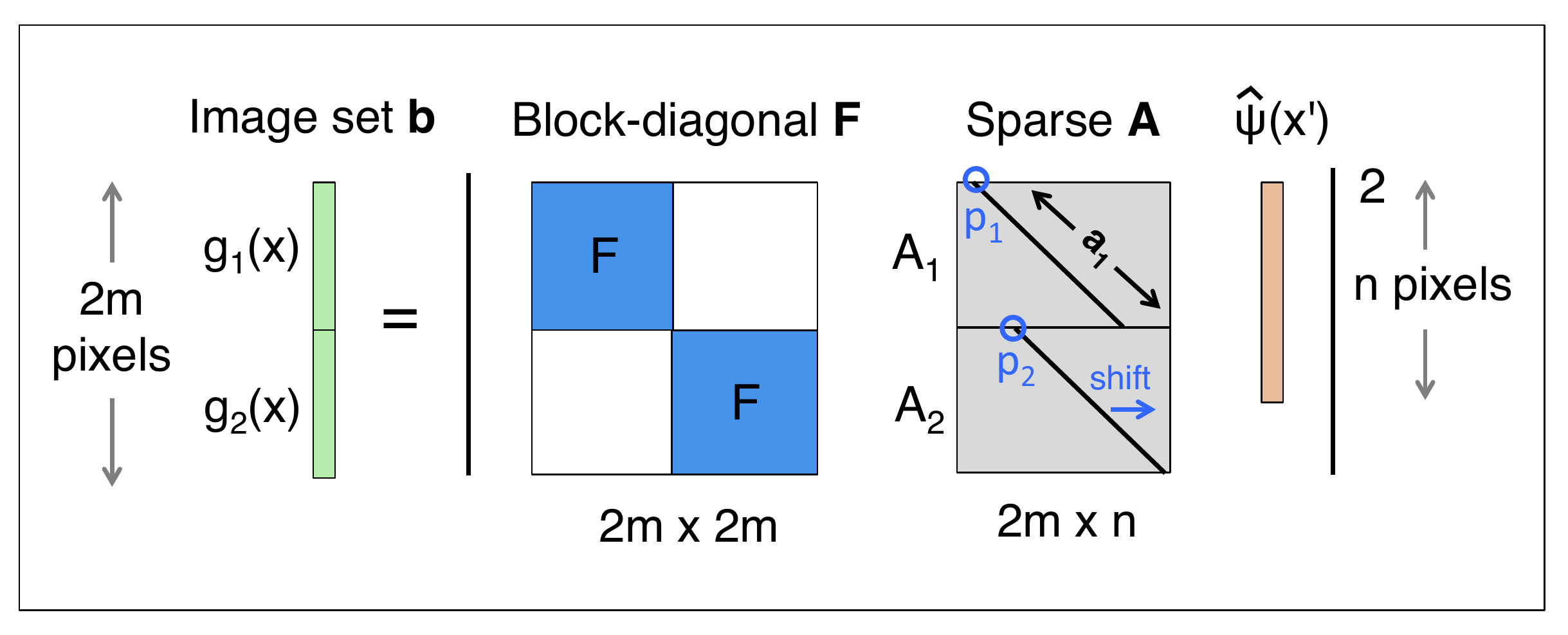}}
\caption{A set of images captured by Fourier ptychography stack together into a long data vector, $\mathbf{b}$. Each associated matrix transform is similarly stacked and combined to form our final measurement matrix, $\mathbf{D}=\mathbf{F}\mathbf{A}$. Here, we show stacking of just two images for simplicity. Typically, over 200 images are stacked. \label{matrix}}
\end{figure}

Fourier ptychography acquires a series of $q$ images, $\{\mathbf{g}_j\}_{j=1}^q$. We combine this image set into a single vector by ``stacking" all images in Eq.\ \ref{image_matrix_2}:
\begin{equation}
\mathbf{b}  = \left|\mathbf{F}\mathbf{A}\hat{\psi}\right|^{2} = \left|\mathbf{D} \hat{\psi}\right|^{2}.
\label{image_matrix_3}
\end{equation}
Here, $\mathbf{b}$ is $\{\mathbf{g}\}$ expressed as a $(q\cdot m \times 1)$ stacked image vector (see Fig.\ \ref{matrix}). In addition, we define $\mathbf{D}=\mathbf{FA}$, where $\mathbf{F}$ is a $(q\cdot m\times q\cdot m)$ block diagonal matrix containing $q$ copies of the low-resolution DFT matrices $\mathbf{F}^{(m)}$ in its diagonal blocks, and $\mathbf{A}$ has size $(q\cdot m \times n)$ and is formed by vertically stacking each aperture matrix $\mathbf{A}_{j}$:
\begin{equation}
\mathbf{F} = \left( \begin{array}
{ccc} \mathbf{F}^{(m)} & \cdots & 0\\ 
\vdots & \ddots & \vdots \\ 
0 & \cdots & \mathbf{F}^{(m)}
\end{array} \right), \;
\mathbf{A}=\left( \begin{array}
{c} \mathbf{A}_1\\ 
\vdots \\ 
\mathbf{A}_q
\end{array} \right).
\label{F_matrix}
\end{equation}
We denote the transpose of the $i$th row of $\mathbf{D}$ as $\mathbf{d}_i$, which is a column vector. The set $\{\mathbf{d}_i\}$ forms our measurement vectors. The measured intensity in the $i$th pixel is the square of the inner product between $\mathbf{d}_i$ and the spectrum $\hat{\psi}$: $b_i = \smash{|\left<\mathbf{d}_{i},\hat{\psi}\right>|}^{2}$. Next, we ``lift'' the solution $\hat{\psi}$ out of the quadratic relationship in Eq.\ \ref{image_matrix_3}. As suggested in \cite{Balan09}, we may instead express it in the space of $(n\times n)$ positive-semidefinite matrices:
\begin{equation}
b_i = \mathrm{Tr}\left(\hat{\psi}^{*}\mathbf{d}_{i}\mathbf{d}_{i}^{*}\hat{\psi}\right) =\mathrm{Tr}\left(\mathbf{d}_{i}\mathbf{d}_{i}^{*}\hat{\psi} \hat{\psi}^{*}\right)
=\mathrm{Tr}\left(\mathbf{D}_{i}\mathbf{X}\right),
\label{lifting}
\end{equation}
where $\mathbf{D}_{i}=\mathbf{d}_{i}\mathbf{d}_{i}^{*}$ is a rank-1 measurement matrix constructed from the $i$th measurement vector $\mathbf{d}_i$, $\mathbf{X} = \hat{\psi}\hat{\psi}^{*}$ is an $(n\times n)$ rank-$1$ outer product, and $1 \le i \le q \cdot m$. Equation \ref{lifting} states that our quadratic image measurements $\{b_i\}_{i=1}^{q\cdot m}$ are linear transforms of $\hat{\psi}$ in a higher dimensional space. We may combine these $q\cdot m$ linear transforms into a single linear operator $\mathscr{A}$ to summarize the relationship between the stacked image vector $\mathbf{b}$ and the matrix $\mathbf{X}$ as, $\mathscr{A}(\mathbf{X}) = \mathbf{b}$.

One can now pose the phase retrieval problem as a rank minimization procedure:
\begin{equation}
\begin{aligned}
& \text{minimize}
& & \mathrm{rank}(\mathbf{X}) \\
& \text{subject to}
& & \mathscr{A}(\mathbf{X}) = \mathbf{b} , \\
&&& \mathbf{X} \succeq 0,
\end{aligned}
\label{rank_min}
\end{equation}  
where $\mathbf{X} \succeq 0$ denotes $\mathbf{X}$ is positive-semidefinite. This rank minimization problem is not convex and is a computational challenge. Instead, adapting ideas from \cite{Fazel02}, we form a convex relaxation of Eq.\ \ref{rank_min} by replacing the rank of matrix $\mathbf{X}$ with its trace. This creates a convex semidefinite program:
\begin{equation}
\begin{aligned}
& \text{minimize}
& & \mathrm{Tr}(\mathbf{X}) \\
& \text{subject to}
& &  \mathscr{A}(\mathbf{X}) = \mathbf{b} , \\
&&& \mathbf{X} \succeq 0.
\end{aligned}
\label{trace_min}
\end{equation}  
Several recent results establish that the relaxation in Eq.\ \ref{trace_min} is equivalent to Eq.\ \ref{rank_min} under certain conditions on the operator $\mathscr{A}$~\cite{Recht07,Candes13}. Although not necessarily equivalent in general, this relaxation consistently offers us highly accurate experimental performance. To account for the presence of noise, we may reform Eq.\ \ref{trace_min} such that the measured intensities in $\mathbf{b}$ are no longer strictly enforced constraints, but instead appear in the objective function:
\begin{equation}
\begin{aligned}
& \text{minimize}
& & \alpha \thinspace \mathrm{Tr}(\mathbf{X}) + \frac{1}{2} \|\mathscr{A}(\mathbf{X})-\mathbf{b}\|\\
& \text{subject to}
& & \mathbf{X} \succeq 0.
\end{aligned}
\label{trace_min_final}
\end{equation}
Here, $\alpha$ is a scalar regularization variable that directly trades off goodness for complexity of fit. Its optimal value depends upon the assumed noise level. Equation \ref{trace_min_final} forms our final convex problem to recover a resolution-improved complex sample $\psi$ from a set of obliquely illuminated images in $\mathbf{b}$. Many standard tools are available to solve this semidefinite program (see Methods section). Its solution defines our Convex Lifted Ptychography (CLP) approach. 

In practice, CLP returns a low-rank matrix $\mathbf{X}$, with a rapidly decaying spectrum, as the optimal solution of Eq.\ \ref{trace_min_final}. We obtain our final complex image estimate $\psi$ by first performing a singular value decomposition of $\mathbf{X}$. Given low-noise imaging conditions and spatially coherent illumination, we set $\psi$ to the Fourier transform of the largest resulting singular vector. Viewed as an autocorrelation matrix, we may also find useful statistical measurements within the remaining smaller singular vectors of $\mathbf{X}$. We note that one may also identify $\mathbf{X}$ as the discrete mutual intensity matrix of a partially coherent optical field~\cite{Ozaktas02}, and view Eq.\ \ref{trace_min_final} as an alternative solver for the stationary mixed states of a ptychography setup~\cite{Thibault13}.

Without any further modification, three points distinguish Eq.\ \ref{trace_min_final} from existing AP-based ptychography solvers. First, the convex solver has a larger search space. If AP is used to iteratively update an $n$-pixel estimate, Eq.\ \ref{trace_min_final} must solve for an $n\times n$ positive-semidefinite matrix. Second, this boost in the solution space dimension guarantees the convex program may find its global optimum with tractable computation. This allows CLP to avoid AP's frequent convergence to local minima (i.e., failure to approach the true image). Finally, Eq.\ \ref{trace_min_final} implicitly considers the presence of noise by offering a parameter ($\alpha$) to tune with an assumed noise level. AP-based solvers lack this parameter and can be easily led into incorrect local minima by even low noise levels, which we demonstrate next.

\begin{figure}
\centerline{\includegraphics[width=.45\textwidth]{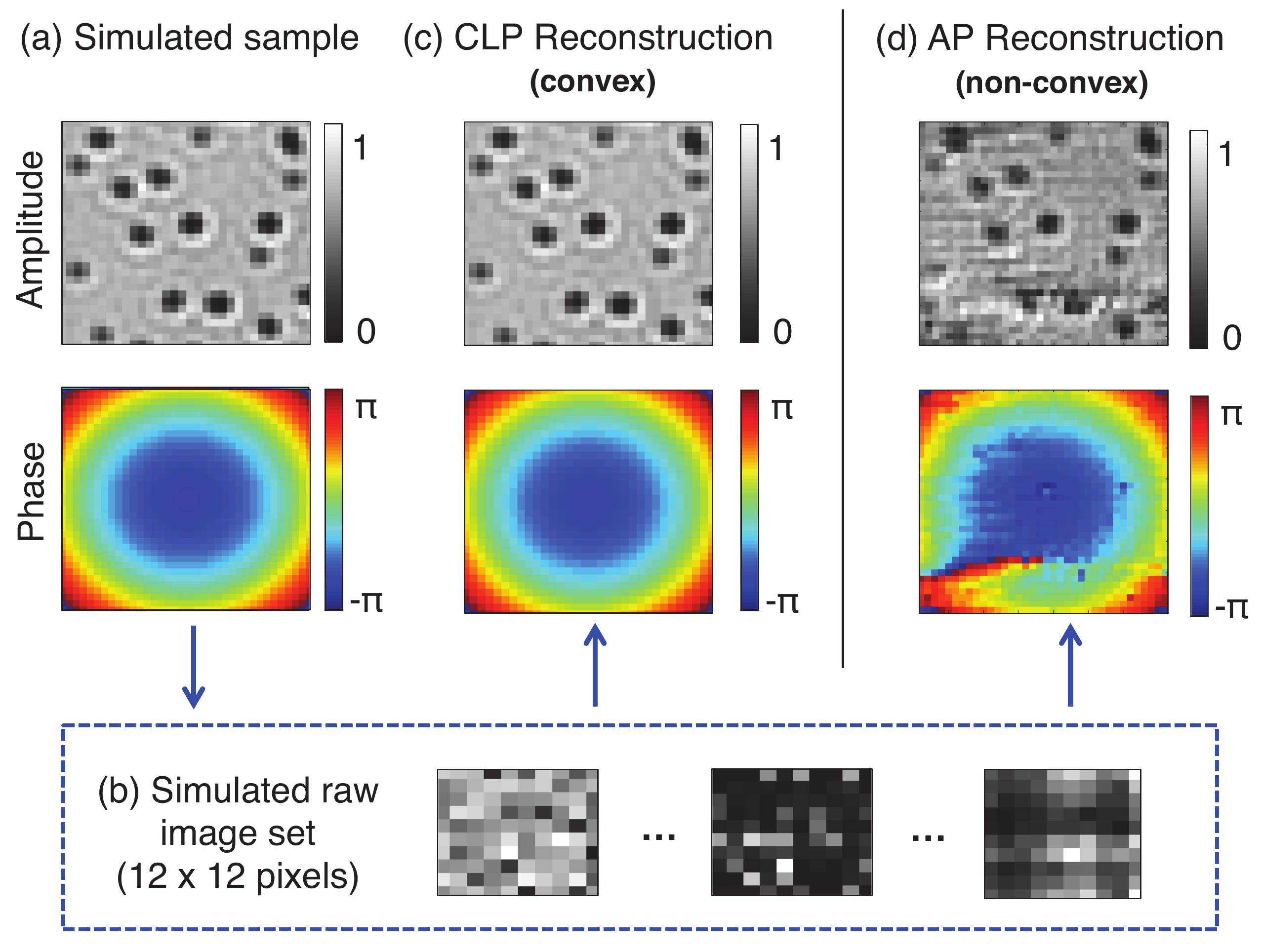}}
\caption{Simulation of the CLP algorithm. (a) An $n=36\times 36$ pixel complex sample (simulated) consisting of microspheres modulated with an independent quadratic phase envelope. (b) Sequence of low-resolution simulated intensity measurements ($m=12\times 12$ pixels each), serving as algorithm input. (c)-(d) Example CLP and AP reconstructions, where CLP is successful but AP converges to an incorrect local minimum. Here we use $q=8^{2}$ images to achieve a resolution gain of 3 along each spatial dimension and simultaneously acquire phase. \label{CLP_sim}}
\end{figure}

\subsection{CLP simulations and noise performance}

We simulate Fourier ptychography following the setup in Fig.\ \ref{setup}. We capture multiple two-dimensional images in $(x,y)$ from a three-dimensional optical geometry. The simulated FP setup contains a detector with $m=12^{2}$ pixels that are each $4\mu$m wide, a 0.1 NA lens ($6^{\circ}$ collection angle) with unity magnification, and an LED array designed to offer an illumination NA of 0.2 ($\theta_{max}=11.5^\circ$ maximum illumination angle). These parameters define the reconstructed resolution of our complex sample as $n=36^{2}$ pixels, increasing the pixel count of one raw image by a factor $n/m=9$.

Fig.\ \ref{CLP_sim}(b) shows example simulated raw images from a sample of absorptive beads modulated by a quadratic phase envelope. Within each raw image, the set of absorptive beads is not clearly resolved. Here, we simulate the capture of $q=8^{2}$ low resolution images, each uniquely illuminated from one of $q=8^{2}$ LEDs in the square array. We then input this image set into both the standard AP algorithm~\cite{Faulkner04} as well as CLP in Eq.\ \ref{trace_min_final} (setting $\alpha=.001$) to recover a high resolution ($36\times36$ pixel) complex sample. Even in the noiseless case, 5 iterations of nonlinear AP introduces unpredictable artifacts to both the recovered amplitude and phase (Fig.\ \ref{CLP_sim}(d)), while CLP offers near perfect recovery (Fig.\ \ref{CLP_sim}(c)). A constant phase offset is subtracted from both reconstructions for fair comparison.

Next, we quantify AP and CLP performance. We repeat the reconstructions in Fig.\ \ref{CLP_sim}, fixing $\alpha=.001$ in Eq.~\ref{trace_min_final} while varying two relevant parameters: the number of captured images $q$, and their signal-to-noise ratio (SNR). We define the SNR as, $\textrm{SNR} = 10\log_{10}(\left<|\psi|^{2}\right>/\left<|N^{2}|\right>)$, where $\left<|\psi|^{2}\right>$ is the mean sample intensity and $\left<|N^{2}|\right>$ is the mean intensity of uniform Gaussian noise added to each simulated raw image. To account for the unknown constant phase offset in all phase retrieval reconstructions, we follow \cite{Maiden09} and define our reconstruction mean-squared error as $\mathrm{MSE}=\sum_x \left|\psi(x)-\rho s(x)\right|^{2}/\sum_x \left|\psi(x)\right|^{2}$, where $\rho = \sum_x \psi(x)s^{*}(x)/\sum_x\left|s(x)\right|^{2}$ is constant phase factor shifting our reconstructed phase to optimally match the known phase of the ground truth sample. 

\begin{figure} [t]
\centerline{\includegraphics[width=.45\textwidth]{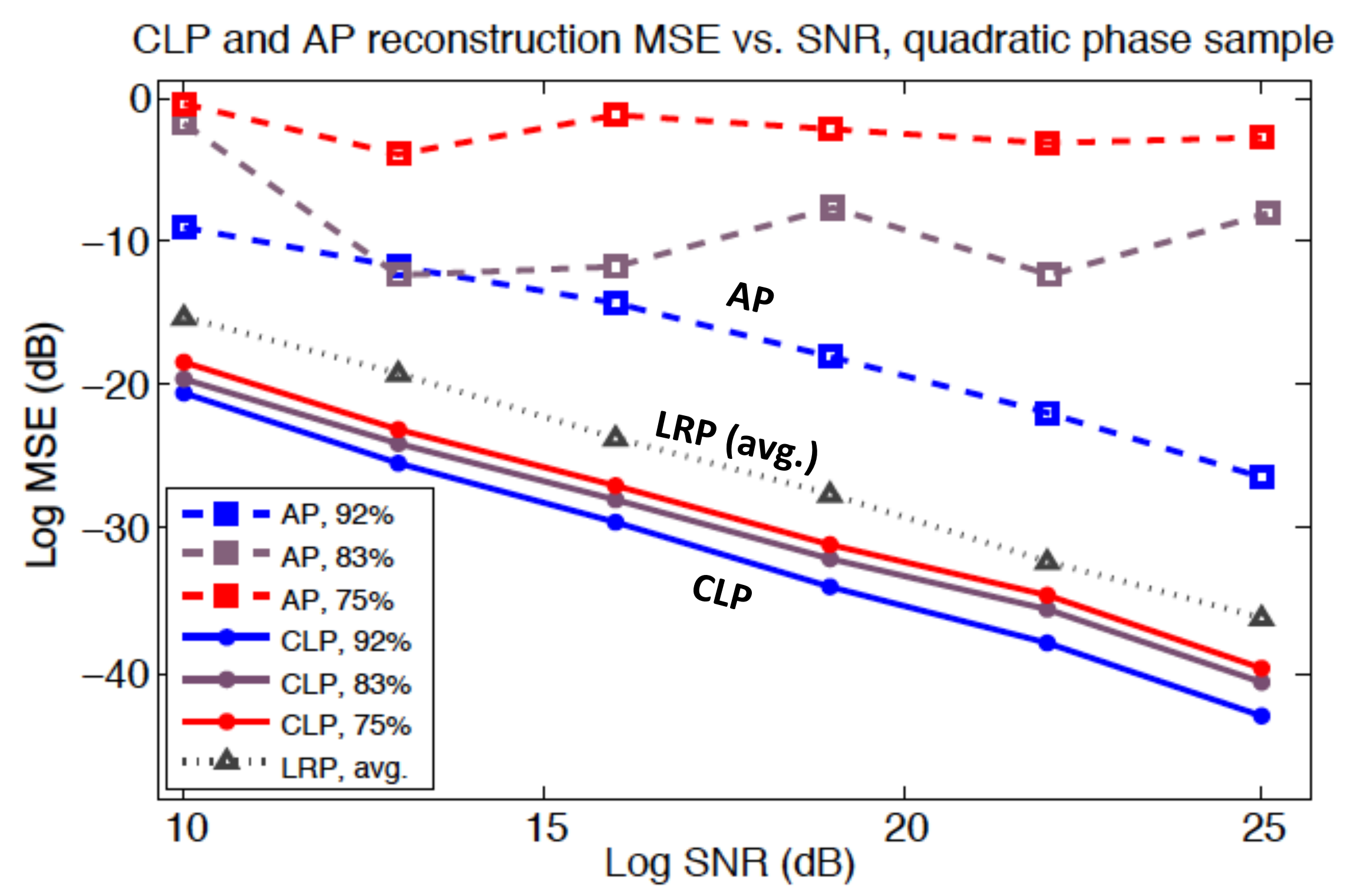}}
\caption{Reconstruction MSE versus signal to noise ratio (SNR) of CLP and AP (log scale, dB). Each curve represents reconstruction with a different number of captured images, $q$, corresponding to a different percentage of spectrum overlap ($ol$, noted in legend). Each point is an average over 5 independent algorithm runs with unique noise. Also included is the average performance of our LRP algorithm over the same 3 spectrum overlap settings.\label{snr_plots}}
\end{figure}

Figure \ref{snr_plots} plots MSE as a function of SNR for this large set of CLP and AP reconstructions. Each of the algorithms' 3 independent curves simulates reconstruction using a different number of captured images, $q$. We summarize $q$ by defining a Fourier spectrum overlap percentage: $ol=1-(n-m)/qm$. Each of the 6 points within one curve simulates a different level of measurement noise. Each point is an average over 5 independent trials. Since AP tends not to converge in the presence of noise, we represent each AP trial with the reconstruction that offers the lowest MSE across all iteration steps (up to 20 iterations). All CLP reconstructions improve linearly as SNR increases, while AP performance fluctuates unpredictably. For both algorithms, performance improves with increased spectrum overlap $ol$, and reconstruction fidelity quickly deteriorates and then effectively fails when $ol$ drops below ${\raise.17ex\hbox{$\scriptstyle\sim$}}60\%$.

\section{Low-rank factorization}

Posing the inverse problem of ptychography as a semidefinite program (Eq.\ \ref{trace_min_final}) is a good first step towards a more tractable solver. However, its constraint that $\mathbf{X}$ remain positive-semidefinite is computationally demanding: each iteration typically requires a full eigenvalue decomposition of $\mathbf{X}$. As the size of $\mathbf{X}$ scales with $n^2$, processable image sizes are limited to an order of $10^{4}$ pixels on current desktop machines. This scaling limit does not prevent large-scale CLP processing of ptychography data. It is common practice to segment each detected image into as few as $10^{3}$ pixels, process each segment separately, and then ``tile" the resulting reconstructions back together into a final full-resolution algorithm output~\cite{Zheng13}. CLP may also parallelize its computation with this strategy.

While such tiling parallelization offers significant speedup, a simple observation helps avoid the poor scaling of CLP altogether: the desired solution of the ptychography problem in Eq.\ \ref{rank_min} is low-rank. Instead of solving for an $n\times n$ matrix $\mathbf{X}$, it is thus natural to adopt a low-rank ansatz and factorize the matrix $\mathbf{X}$ as $\mathbf{X}=\mathbf{R} \mathbf{R}^T$, where our decision variable $\mathbf{R}$ is an $n \times r$ rectangular matrix with $r < n$~\cite{Burer03,Burer05}. Inserting this factorization into our optimization problem in Eq.\ \ref{trace_min} and writing the constraints in terms of the measurement matrix  $\mathbf{D}_i = \mathbf{d}_i \mathbf{d}_i^T$ creates the non-convex program,
\begin{equation}
\begin{aligned}
& \text{minimize}
& & \mathrm{Tr}(\mathbf{R}\mathbf{R}^T) \\
& \text{subject to}
& & \mathrm{Tr}(\mathbf{D}_i \mathbf{R}\mathbf{R}^T) = b_i \quad \text{for all $i$}.
\end{aligned}
\label{trace_min_factored}
\end{equation}
Besides removing the positive semidefinite constraint in Eq.\ \ref{trace_min}, the factored form of Eq.\ \ref{trace_min_factored} presents two more key adjustments to our original convex formulation. First, using the relationship $\mathrm{Tr}(\mathbf{R}\mathbf{R}^T)=\|\mathbf{R}\|_{F}^{2}$, it is direct to rewrite the objective function and each constraint in Eq.\ \ref{trace_min_factored} with just one $n\times r$ matrix $\mathbf{R}$. Now instead of storing an $n\times n$ matrix like CLP, LRP must only store an $n\times r$ matrix. Since most practical applications of ptychography require coherent optics, the desired solution rank $r$ will typically be close to 1, thus significantly relaxing storage requirements (i.e., LRP memory usage now scales linearly instead of quadratically with n).

Second, the feasible set of Eq.\ \ref{trace_min_factored} is no longer convex. We thus must shift our solution strategy away from a simple semidefinite program. Prior work in \cite{Burer03,Burer05} suggests that an efficient and practically successful method of solving Eq.\ \ref{trace_min_factored} is to minimize the following augmented Lagrangian function:
\begin{equation} \label{eqn:augmented-lagrangian}
\begin{multlined}
 \hspace*{-.3cm}
L(\mathbf{R}, \mathbf{y}, \sigma) = \mathrm{Tr}(\mathbf{R}\mathbf{R}^T) - \sum\nolimits_i y_i \cdot \left( \mathrm{Tr}(\mathbf{D}_i \mathbf{R} \mathbf{R}^T) - b_i \right) \\
+ \frac{\sigma}{2} \cdot \sum\nolimits_i \left( \mathrm{Tr}(\mathbf{D}_i \mathbf{R} \mathbf{R}^T) - b_i \right)^2
\end{multlined}
\end{equation}
where $\mathbf{R} \in \mathbb{R}^{n \times r}$ is the unknown decision variable and the two variables $y \in \mathbb{R}^{q\cdot m}$ and $\sigma \in  \mathbb{R}^{+}$ are new parameters to help guide our algorithm to its final reconstruction. The first term in Eq.\ \ref{eqn:augmented-lagrangian} is the objective function from Eq.\ \ref{trace_min_factored}, indirectly encouraging a low-rank factorization product. This tracks our original assumption of a rank-1 solution within a ``lifted" solution space. The second term contains the known equality constraints in Eq.\ \ref{trace_min_factored} (i.e., the measured intensities), each assigned a weight $y_i$. The third term is a penalized fitting error that we abbreviate with label $v$. It is weighted by one penalty parameter $\sigma$, mimicking the role of a Lagrangian multiplier. With an appropriate fixed selection of $y_i$'s and $\sigma$, the minimization of $L(\mathbf{R}, \mathbf{y}, \sigma)$ with respect to $\mathbf{R}$ will produce our desired solution. As an unconstrained function, the minimum of $L$ is quickly found via standard tools (e.g., a quasi-Newton approach such as the LBFGS algorithm~\cite{Schmidt14}).


The goal of our low-rank ptychography (LRP) algorithm may now be stated as follows: determine a suitable set of $(y_i,\sigma)$ such that Eq.\ \ref{eqn:augmented-lagrangian} may be minimized with respect to $\mathbf{R}$ to find the optimal solution. We use the iterative algorithm suggested in \cite{Burer03} to sequentially minimize $L$ with respect to $\mathbf{R}^k$ at iteration $k$, and then update a new parameter set $(y^{k+1},\sigma^{k+1})$ at iteration $k+1$. We update parameters $(y^{k+1},\sigma^{k+1})$ to ensure their associated term's contribution to the summation forming $L$ is relatively small. This suggests $\mathbf{R}^{k+1}$ is proceeding to a more feasible solution. The relative permissible size of the second and third terms in $L$ are controlled by two important parameters, $\eta<1$ and $\gamma>1$: if the third term $v$ sufficiently decreases such that $v^{k+1} \le \eta v^k$, then we hold its multiplier $\sigma$ fixed and update the equality constraint multipliers, $y_i$. Otherwise, we increase $\sigma$ by a factor $\gamma$ such that $\sigma^{k+1}=\gamma\sigma^k$. A detailed discussion of these algorithmic steps is in \cite{Burer03,Burer05}.

We initialize the LRP algorithm with an estimate of the unknown high-resolution complex sample function $\psi_0$, contained within a low-rank matrix $\mathbf{R}^0$. We terminate the algorithm either if it reaches a sufficient number of iterations, or if the minimizer fulfills some convergence criterion. We form $\mathbf{R}^0$ using a spectral method, which can help increase solver accuracy and decrease computation time~\cite{Candes14}. Specifically, we select the $r$ columns of $\mathbf{R}^0$ as the leading $r$ eigenvectors of $\mathbf{D}^{*}\textrm{diag}[\textbf{b}]\mathbf{D}$, where $\mathbf{D}$ is the measurement matrix in Eq.\ \ref{image_matrix_3}. While this spectral approach works quite well in practice, a random initialization of $\mathbf{R}^0$ also often produces an accurate reconstruction.     

\vspace{-.7cm}
\subsection{LRP simulations and noise performance}

\begin{figure}[t]
\centerline{\includegraphics[width=.42\textwidth]{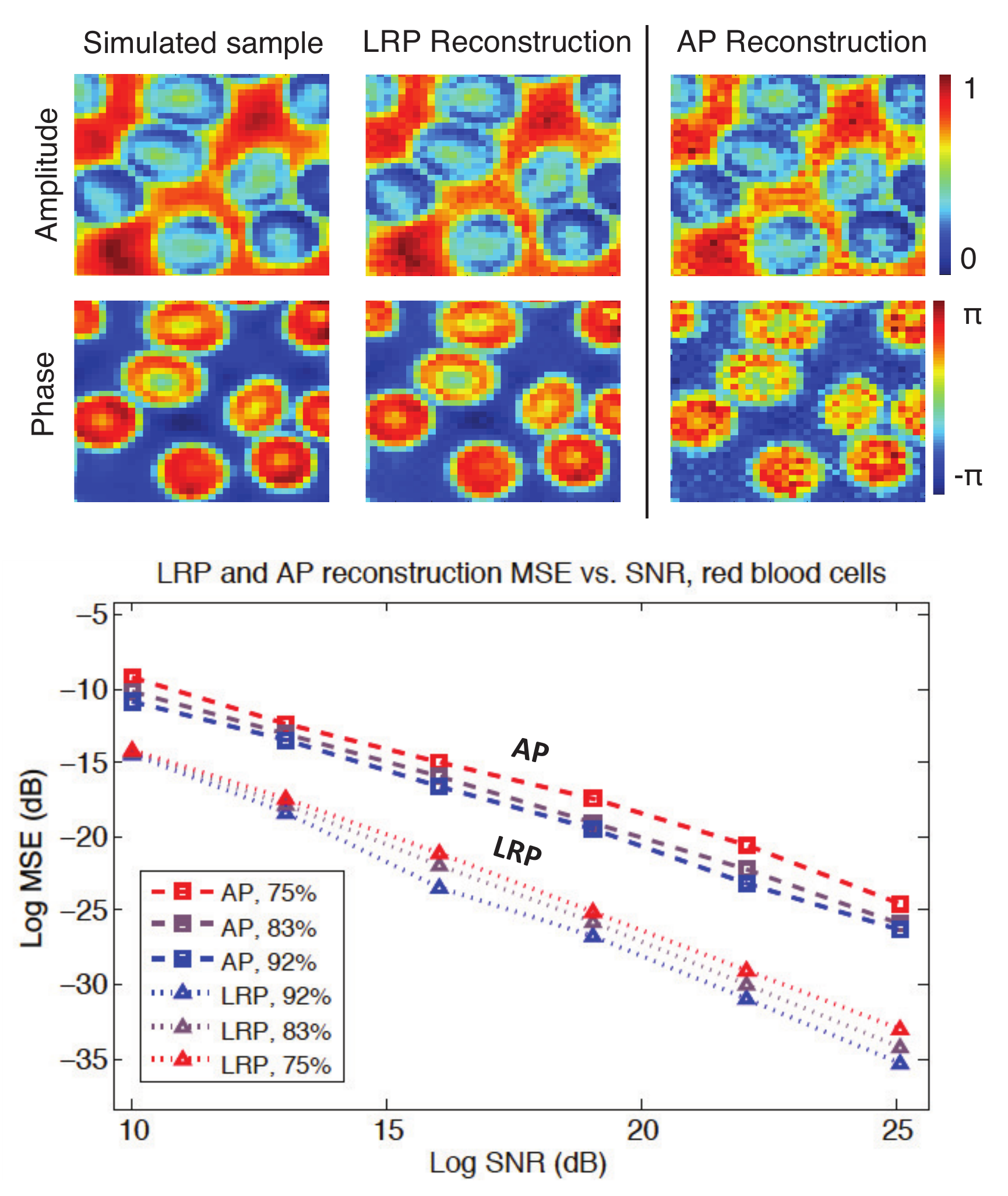}}
\caption{Simulation of the LRP algorithm using the same parameters as for Fig.~\ref{CLP_sim}--Fig.~\ref{snr_plots}, but now with a different ``red blood cell" sample. (Top) From $8^{2}$ noisy raw images (SNR=19), both LRP and AP successfully recover the primary features of this simulated biological sample. (Bottom) MSE versus SNR plot for the red blood cells with simulated noise. The MSE for LRP is {\raise.17ex\hbox{$\scriptstyle\mathtt{\sim}$}}5-10 dB lower than for AP across all noise levels and aperture overlap settings (each point from 5 independent trials). \label{LRP_sim}}
\end{figure}

Following the same procedure used to simulate the CLP algorithm, we test the MSE performance of the LRP algorithm as a function of SNR in Fig.\ \ref{LRP_sim}. We again add different amounts of uncorrelated Gaussian noise to each simulated raw image set and compare the LRP reconstruction with ground truth. This simulated sample is the experimentally obtained amplitude and phase of a human blood smear. It is qualitatively similar to the sample used in Fig.\ \ref{CLP_sim}. Unlike with the simulations in Fig.\ \ref{CLP_sim}--\ref{snr_plots}, the AP algorithm no longer malfunctions at lower spectrum overlap percentages (i.e., lower values of $ol$). Despite this apparent success, the AP minimizer MSE is still an average 5 dB worse than the LRP minimizer MSE across all levels of SNR, achieved without additional parameter optimization or explicit noise modeling. 

In these simulations, we somewhat arbitrarily fix $\eta$ and $\gamma$ at 0.5 and 1.5, respectively, and set the desired rank of the solution, $r$, to 1. In practice, these free variables offer significant freedom to tune the response of LRP to noise. For example, similar to the noise parameter $\alpha$ in Eq.\ \ref{trace_min_final}, the multiplier $\sigma$ (controlled via $\gamma$) in Eq.\ \ref{eqn:augmented-lagrangian} helps trade off complexity for goodness of fit by re-weighting the quadratic fitting error term.

In addition to reducing required memory, the LRP algorithm also improves upon the computational cost of CLP. For an $n$-pixel sample reconstruction, the per iteration cost of the CLP algorithm is currently $O(n^3)$, using big-$O$ notation. The positive-semidefinite constraint in Eq.\ \ref{trace_min_final}, which requires a full eigenvalue decomposition, defines this behavior limit. The per-iteration cost of the LRP algorithm, on the other hand, is $O(n\log{n})$. This large per-iteration cost reduction is the primary source of LRP speedup. For example, LRP required  {\raise.17ex\hbox{$\scriptstyle\sim$}}21 seconds to complete an average simulation of the example in Fig.\ \ref{CLP_sim}, while CLP required {\raise.17ex\hbox{$\scriptstyle\sim$}}170 minutes and AP 1 second on the same desktop machine.

\section{Experimental results}

We experimentally verify the accuracy and improved noise stability of LRP using a Fourier ptychographic microscope, which captures a sequence of images under unique LED illumination following the protocol in \cite{Zheng13} (see Methods section). We first quantitatively verify that LRP accurately measures high resolution and sample phase. This verification highlights how LRP avoids certain undesirable artifacts commonly found in AP reconstructions. Second, we qualitatively compare the high-resolution AP and LRP reconstructions of a biological sample. This set of experiments demonstrates the improved noise stability of our LRP solver.

\subsection{Quantitative performance}

To quantitatively verify resolution improvement, we turn on each of $15\times 15$ LEDs in an array placed 80 mm beneath a calibration target (U.S. Air Force [USAF] resolution chart), capturing a total of 225 low-resolution images. Each LED has an approximate 20 nm spectral bandwidth centered at $\lambda=632$ nm. Using a 0.08 NA microscope objective ($5^{\circ}$ collection angle) and a 0.35 illumination NA ($\theta_{max}=20^{\circ}$ illumination angle), this setup is designed to offer a total complex field resolution gain of $n/m = 25$. Each image spectrum overlaps by $ol\approx70\%$ in area with each neighboring image spectrum. 

\begin{figure}[t]
\centerline{\includegraphics[width=.46\textwidth]{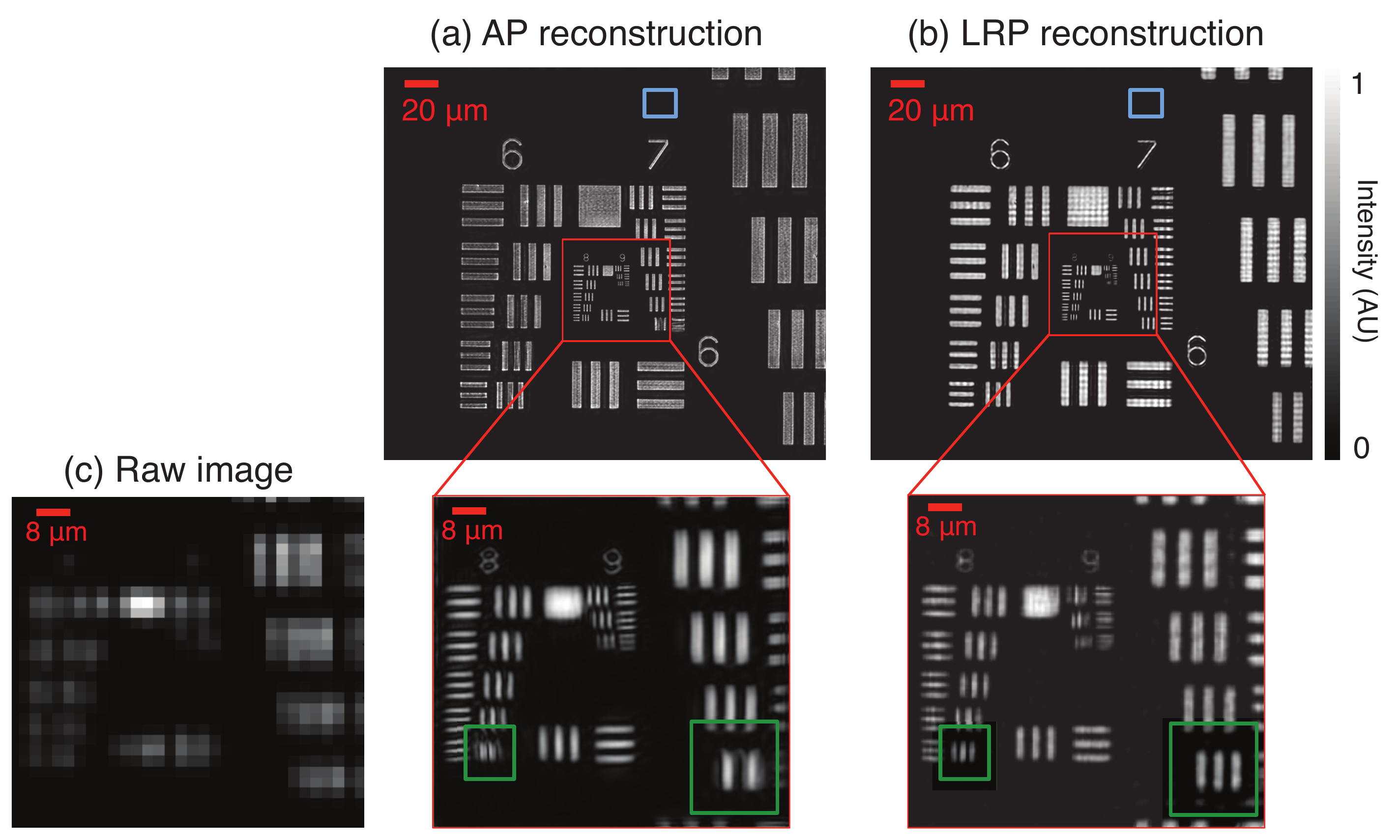}}
\vspace{5 mm}
\caption{Experimental reconstruction of a USAF target showing 5X resolution improvement along both $x$ and $y$ using two ptychography algorithms: (a) AP and (b) LRP. Here we only show reconstructed intensity. LRP avoids artifacts (e.g., boxed in green) commonly encountered in the AP approach. Variances measured in blue boxes. (c) Same cropped region of one low-resolution raw image, for comparison. \label{exper1}}
\end{figure}

For reconstruction, we select $n=25\cdot m$ and use the same aperture parameters with AP and LRP to create the high-resolution images in Fig.\ \ref{exper1} (details in Methods section). Both {\raise.17ex\hbox{$\scriptstyle\sim$}}1 megapixel reconstructions achieve their maximum expected resolving power (Group 9, Element 3: 1.56 $\mu$m line pair spacing). This is approximately 5 times sharper than the smallest resolved feature in one raw image (Group 7, Element 2 in Fig~\ref{exper1}(c)). Our new LRP algorithm avoids certain artifacts that are commonly observed during the nonlinear descent of AP (boxed in green). Both reconstructions slowly fluctuate across areas in the background that we expect to be uniformly bright or dark. These fluctuations are caused in part by imperfect estimates of the aperture function $a$ and LED shift values $p_j$. In a representative background area marked in Fig.\ \ref{exper1} by a $40^{2}$ pixel blue box, AP and LRP exhibit normalized background amplitude variances of $\sigma_A^2=5.4\times10^{-4}$ and $\sigma_L^2=5.0\times10^{-4}$, respectively. Accounting for experimental uncertainty in the values of $a$ and $p_j$ (e.g., following \cite{Horstmeyer14b,Ou14}) may reduce this error. 

To verify that our LRP solver reconstructs quantitatively accurate phase, we next image a monolayer of polystyrene microspheres (index of refraction $n_{m}=1.587$) immersed in oil ($n_{o}=1.515$, both indexes for $\lambda=632$ nm light). To demonstrate the LRP algorithm easily generalizes to any ptychographic arrangement, we perform this experiment on a new FP microscope using a larger 0.5 NA objective lens (collection angle = $30^{\circ}$). We now turn on each of 29 LEDs in an array placed 40 mm beneath the sample (0.7 illumination NA, $\theta_{max}=45^{\circ}$ illumination angle), capturing 29 unique images. The synthesized numerical aperture of this new FP microscope, computed as the sum of the illumination NA and objective lens NA, is NA$_{s}=1.2$. With a greater-than-unity synthetic NA, our reconstructions should offer oil-immersion quality resolution ({\raise.17ex\hbox{$\scriptstyle\mathtt{\sim}$}}385 nm smallest resolvable feature spacing), without requiring any immersion medium between the sample and objective lens.

\begin{figure}[b]
\centerline{\includegraphics[width=.49\textwidth]{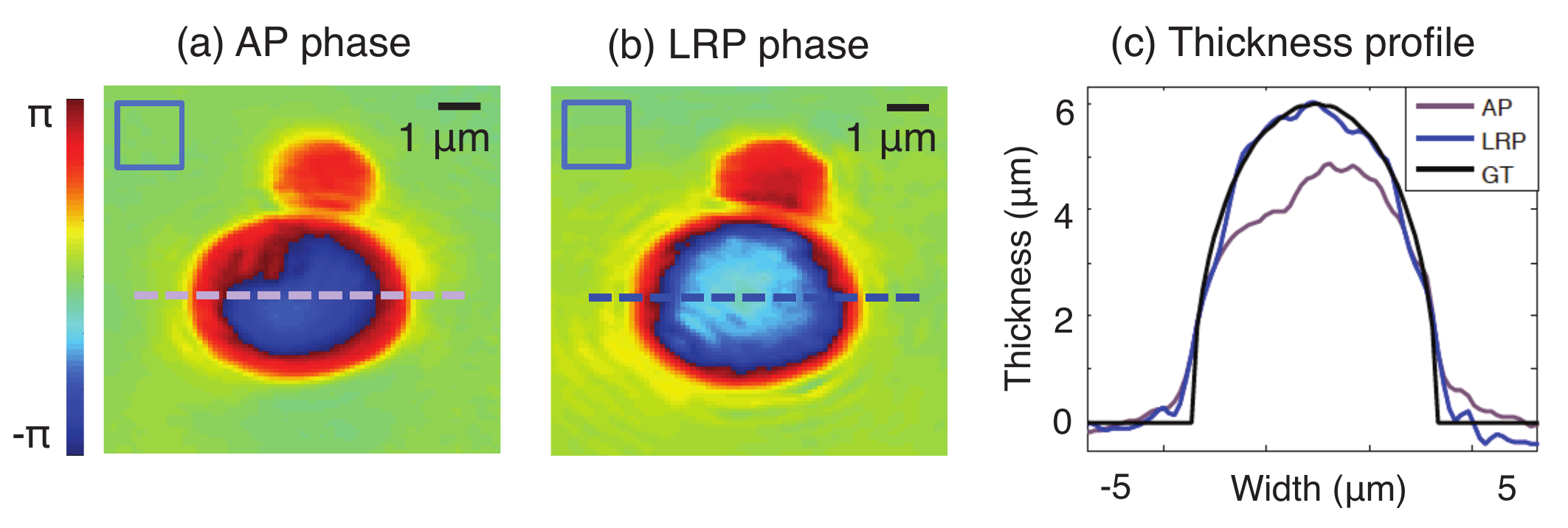}}
\vspace{5 mm}   
\caption{Experimental demonstration of quantitative phase imaging of two polystyrene microspheres. Both (a) AP and (b) LRP reconstruct phase maps that appear qualitatively similar, although the AP phase map flattens at the sphere's center. Variances measured in blue boxes. (c) Plot of microsphere thickness from a trace through the center of the large sphere (dashed line) demonstrates close agreement between LRP and ground truth (GT). 
 \label{exper2}}
\end{figure}

Using the same data and parameters for AP and LRP input, we obtain the high-resolution phase reconstructions of two adjacent microspheres (diameters 3 $\mu$m and 6 $\mu$m) in Fig.\ \ref{exper2}. We have subtracted a constant phase offset from the LRP solution in (b) to allow for direct comparison to the AP solution in (a). The two reconstructions appear qualitatively similar except at the center of the 6 $\mu$m sphere, where the AP phase profile unexpectedly flattens. We highlight this flattening by selecting phase values along each marked dashed line to plot the resulting sample thickness in Fig.\ \ref{exper2}(c). Phase $\phi$ and sample thickness $t$ are related via $t=k\Delta\phi(n_{m}-n_{o})^{-1}$, where $k$ is the average wavenumber and $\Delta\phi=\phi-\phi_0$ is the reconstructed phase minus a constant offset. LRP closely matches the optical thickness of a ground-truth sphere (GT, black curve): the length of the vertical chord connecting the top and bottom arcs of a 6 $\mu$m diameter circle. The normalized amplitude variances from a $40^{2}$-pixel background area are $\sigma_A^2=9.2\times10^{-4}$ and $\sigma_L^2=5.8\times10^{-4}$, respectively, again supporting our observation that LRP reconstructions are more accurate than AP reconstructions. 


\subsection{Biological sample reconstruction}

Our third imaging example uses the same high-NA FP configuration (collection angle = $30^{\circ}$, $\theta_{max}=45^{\circ}$) to resolve a biological phenomenon: the infectious spread of malaria in human blood. The early stages of a \emph{Plasmodium falciparum} infection in erythrocytes (i.e., red blood cells) includes the formation of small parasitic ``rings". It is challenging to resolve these parasitic rings under a microscope without using a high-NA lens and an immersion medium, even after appropriate staining (here, a modified Giemsa stain). Oil-immersion is required for an accurate diagnosis of infection~\cite{WHO09}.

We use FP to resolve \emph{Plasmodium falciparum}-infected cells with a 0.5 NA objective lens and no oil in Fig.\ \ref{exper3}. We capture 29 uniquely illuminated images of a sample region containing several infected cells, with the same FP microscope and LED array used for Fig.\ \ref{exper2} (sample preparation details in Methods section). Figure \ref{exper3}(c) contains an example normally illuminated raw image from this setup, which does not clearly resolve the parasite infection. For comparison, we capture the same sample region with a conventional high-NA oil-immersion microscope (NA = 1.25) using Kohler illumination, which clearly resolves parasites within two cells (Fig.\ \ref{exper3}(a)).

Figure \ref{exper3}(d) presents our phase retrieval reconstructions using the standard AP algorithm (sharpest solutions after 6 iterations). Reconstructions from three data sets are included: images captured with a 1 second exposure (top), a 0.25 second exposure (middle), and 0.1 second exposure (bottom). A shorter exposure implies increased noise within each raw image. While the 1 second exposure-based reconstruction resolves each parasite, blurred cell boundaries and non-uniform fluctuations in amplitude suggest an inaccurate AP convergence. 

\setcounter{figure}{7}    
\begin{figure} [h]
\centerline{\includegraphics[width=.49\textwidth]{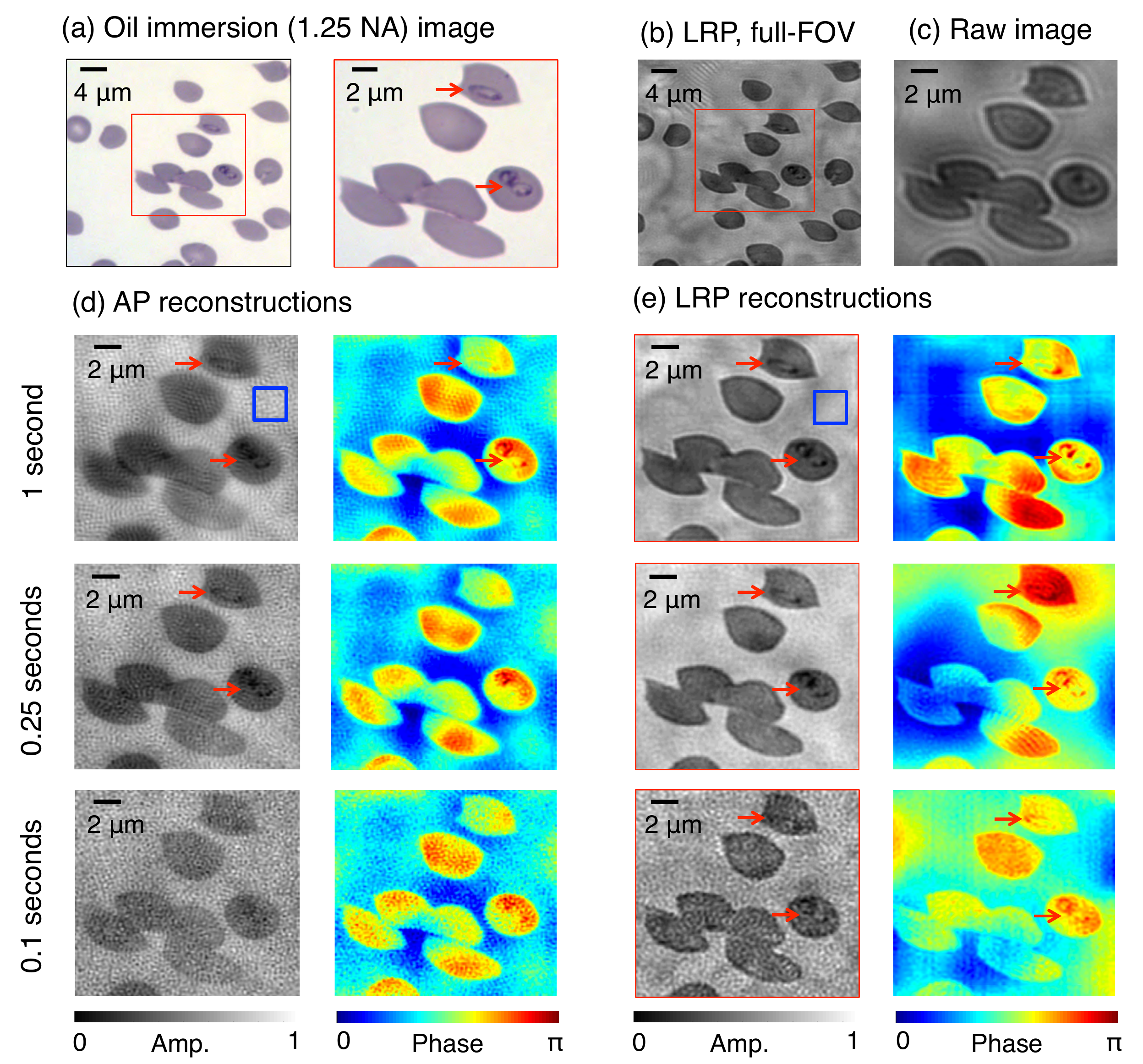}}
\vspace{5 mm}
\caption{Experimental reconstruction of malaria-infected human red blood cells. (a) Oil immersion microscope image (1.25 NA) identifies two infected cells of interest (marked with arrows). (b) Example LRP reconstruction (area of interest in red box). (c) One example raw image used for AP and LRP algorithm input. (d) AP-reconstructed amplitude and phase from three different 29-image data sets, using 1 sec (top), 0.25 sec (middle) and 0.1 sec (bottom) exposure times for all images in each set. Variances measured in blue boxes. Increased noise within short-exposure images deteriorates reconstruction quality until both parasites are not resolved. (e) LRP reconstructions using the same three data sets. Both parasites are clearly resolved in the reconstructed phase for all three exposure levels. \label{exper3}}
\end{figure}
\setcounter{figure}{5}    

The normalized background variance of each AP amplitude reconstruction, from a representative $40^{2}$-pixel window (marked blue square), is $\sigma_A^2=.0020$, $.0027$, and $.0059$ for the 1 sec, 0.25 sec, and 0.1 sec exposure reconstructions, respectively. Both parasite infections are also visible within the 1 sec exposure AP-reconstructed phase. The parasites become challenging to resolve within the phase from 0.25 sec exposure data, and are not resolved within the phase from the 0.1 sec exposure data, due to increased image noise. 

For comparison, reconstructions using our LRP algorithm are shown in Fig.\ \ref{exper3}(e) (sharpest solutions after 15 iterations). For each reconstructed amplitude, we set the desired solution rank to $r=3$. We add the 3 modes of the resulting reconstruction in an intensity basis to create the displayed amplitude images. For each reconstructed phase, we set the desired solution matrix rank to $r=1$ and leave all other parameters unchanged. For all three exposure levels, the amplitude of the cell boundaries remains sharper than in the AP images. 

The normalized amplitude variances from the same background window are now $\sigma_L^2=.0016$ (1 sec), .0022 (0.25 sec), and .0035 (0.1 sec), an average improvement of 26\% from the AP results. Both parasite infections are resolvable in either the reconstructed amplitude or phase, or both, for all three exposure levels. This simple experiment demonstrates that LRP offers a competitive advantage over AP-based algorithms, in the presence of noise, for malaria diagnosis. A shorter image exposure time (i.e., up to 10 times shorter) may still enable accurate diagnosis if LRP is used for ptychographic recovery, which does not appear to be the case when the AP algorithm is used.

\section{Discussion and conclusion}

Through the relaxation in Eq.\ \ref{trace_min}, we first transformed the traditionally nonlinear phase retrieval process for ptychography into a convex problem. We then suggest a practically efficient algorithm to solve the resulting semidefinite program with an appropriate factorization. The result is a new ptychographic image recovery algorithm that is robust to noise. We demonstrate its successful performance in three unique experiments, concluding with a practical biological imaging scenario: the identification of malaria infection without an oil immersion medium and under short-exposure imaging conditions. 

Much future work remains to fully explore the specific benefits of our problem reformulation. Perhaps the most significant departure from prior phase retrieval solvers is the ability to tune the solution rank, $r$. As noted earlier, $r$ connects to statistical features of the optical experiment. We expect an appropriately selected $r$ to enable our solvers to eventually determine the partial coherence of our optical sources, possible vibrations within the setup, sample auto-fluorescence, or even three-dimensional structure. As in prior work with low-rank matrix optimization, we may also artificially enlarge our desired solution rank to encourage the transfer of experimental noise into its smaller singular vectors.

Other extensions of LRP include simultaneously solving for unknown aberrations, systematic setup errors, and inserting additional sample priors such as sparsity. Our connection to the well-understood area of convex optimization will hopefully make the task of analyzing phase retrieval algorithm performance more accessible to the average experimentalist, who may tailor our problem formulation to their own unique application. What's more, we are hopeful that tools from convex analysis will continue to provide useful theoretical guarantees regarding algorithm performance, a crucial feature missing from current nonlinear solvers.


\begin{materials}

\section{The Fourier ptychographic microscope} The FP microscope consists of a 16$\times$16 array of surface-mounted LEDs (model SMD 3528, center wavelength $\lambda$=632 nm, 4 mm LED pitch, 150 $\mu$m active area diameter) placed $l$=40-80 mm beneath the sample plane. Images are recorded with a CCD detector (Kodak KAI-29050) containing 5.5 $\mu$m pixels. The images in Fig.\ \ref{exper1} were captured with a 2X Olympus microscope objective (apochromatic Plan APO 0.08 NA), and we set $l$=80 mm, $m$=48 and $n$=480 for each of the 3$\times$3 image tiles forming our full image (see Algorithm parameters below). For Fig.\ \ref{exper2} and Fig.\ \ref{exper3}, we imaged the microsphere slide and cells with a 20X Olympus microscope objective (0.5 NA UPLFLN), set $l$=42 mm, $m$=160, $n$=320, and used three rings of 8, 8 and 12 evenly spaced LEDs (ring radii=16, 32 and 40 mm, respectively) and one central (on-axis) LED to capture the 29 images. The sample for Fig.\ \ref{exper3} was prepared by maintaining erythrocyte asexual stage cultures of the P. falciparum strain 3D7 in culture medium, then smearing, fixing with methanol and applying a Hema 3 stain (following the protocol in \cite{Hoef13}).

\section{Algorithm parameters} For the reconstructions in Fig.\ \ref{exper1}, we first segmented each raw image into 3$\times$3 tiles (each with $m$=48$^2$ pixels and overlapping with neighboring tiles by 4 pixels). AP and LRP then reconstruct each tile separately before recombining the solutions into the approximate 1 megapixel images in Fig.\ \ref{exper1}. This tiling procedure matches that in \cite{Zheng13}. We determined the optimal number of AP and LRP algorithm iterations as 6 and 15, respectively, and fixed this for each tile here (and all reconstructions in general). LRP's free/initial parameters are typically set to $\gamma$=1.5, $\eta$=0.3, $y^{0}$= and $\sigma^{0}$=10. We determined each aperture function $a$ a-priori using an iterative procedure~\cite{Ou14} and used the same $a$ for each algorithm and each tile. No tiling was used for the reconstructions in Fig.\ \ref{exper2} and Fig.\ \ref{exper3}. For these two figures, we set $a$ to an unmodified circ-function with radius defined by the objective lens NA. Fig.\ \ref{exper2} transforms each $m$=80$^2$ pixel sub-image into its $n$=160$^2$ pixel solutions, while Fig.\ \ref{exper3} transforms each $m$=120$^2$ pixel sub-image into its $n$=240$^2$ pixel solutions. As noted in the main text, a fixed phase background was subtracted from all phase images for fair comparison.

\section{Computational specifics} We performed all processing on a high-end desktop containing two Intel Xeon 2.0 GHz CPUs and two 3GB GeForce GTX GPUs. Code was written in Matlab with built-in GPU acceleration. We solved our CLP semidefinite program using the TFOCS code package~\cite{Becker10}. Our LRP algorithm borrows concepts from the LBFGS solver in \cite{Schmidt14} for one specific minimization step. LRP's total recovery time for the 1 megapixel example in Fig.\ \ref{exper1} was approximately 130 seconds, while AP completed in approximately 15 seconds on the same desktop.

\end{materials}

\begin{acknowledgments}
We thank Stephen Becker for useful suggestions regarding TFOCS, as well as Laura Waller, Lei Tian, Salman Asif and John Bruer for helpful discussions and feedback. R.H., X.O. and C.Y. acknowledge funding support from the National Institutes of Health (grant no.\ 1DP2OD007307-01) and Clearbridge Biophotonics Pte Ltd., Singapore (Agency Award no.\ Clearbridge 1). J.A.T. gratefully acknowledges support from ONR award N00014-11-1002 and a Sloan Research Fellowship. Thanks are also due to the Moore Foundation.
\end{acknowledgments}

\end{article}

%
%
%

\end{document}